\documentstyle[prl,aps,epsfig,twocolumn]{revtex}
\begin{document}
\renewcommand{\thesection}{\Alph{section}}
\title{Duality in semileptonic inclusive $B$-decays in potential models: \\ 
regular versus singular potentials}
\author{A. Le Yaouanc$^{a}$, 
D. Melikhov$^{b}$,\thanks{Alexander-von-Humboldt fellow} 
V. Mor\'enas$^{c}$, L. Oliver$^{a}$, O. P\`ene$^{a}$ and  J.-C. Raynal$^{a}$}
\address{
${}^a$ Laboratoire de Physique Th\'eorique, Universit\'e de Paris XI, 
B\^atiment 210, 91405 Orsay Cedex, France\\
$^b$ Institut f\"ur Theoretische Physik, Universit\"at Heidelberg,
Philosophenweg 16, D-69120, Heidelberg, Germany\\
$^c$ Laboratoire de Physique Corpusculaire, Universit\'e Blaise Pascal - CNRS/IN2P3, 63000
Aubi\`ere Cedex, France} 
\maketitle
\mediumtext
\begin{abstract}
Making use of the nonrelativistic potential model for the description of mesons, and working in  the
Shifman-Voloshin limit, we compare the integrated rate $\Gamma(B\to X_cl\nu)$  calculated as a sum of the
individual decay rates to the quantum-mechanical analog of the OPE. In the case of a potential regular at
the origin, we find a well-defined  duality violation, which is however exponentially  small. It
corresponds  to the charm resonances kinematically forbidden in the decay process, but apparently picked
up by the OPE. For singular potentials, we do not obtain a full OPE series, but only a limited Taylor
expansion, since the cofficients become infinite beyond some order. In this case, we do not find an
indication of duality violation: the difference is smaller than the last term of the limited expansion.
This emphasizes that the case of singular potentials,  which may be relevant for QCD, deserves further
study. 

\vskip.1cm
\noindent PACS number(s): 13.20 He, 12.39 Jh, 12.39.Pn
\end{abstract}

\vskip .6cm

\narrowtext
The theoretical framework based on the Operator Product Expansion (OPE) 
determines in QCD the heavy meson inclusive decay rate as series in inverse 
powers of the heavy quark mass,  
with the coefficients proportional to the meson matrix elements of the local 
operators of increasing dimensions \cite{cgg,bsuv93}. 
The calculation is based on representing the decay rate as the contour integral 
in the complex $q_0$-plane. The OPE makes the  
contour integrals easily calculable term by term and provides the decay rate as a 
$1/m_Q$ series. 

There are however potentially dangerous points in this calculation: 

\noindent
(i) the OPE series is at best assymptotically convergent even for large absolute 
values of the complex $q_0$,

\noindent
(ii) the integration contour for the decay rate contains a segment near the 
physical region, where the OPE cannot be justified \cite{cgg}. 

This might lead to the violation of duality for the decay rate, i.e. to the difference 
between the OPE-calculated decay rate and the result of summing the individual decay rates 
of the opened channels. This issue was also discussed by N. Isgur \cite{isgur}. 

In this letter we discuss the semileptonic decay rate in the small velocity limit 
and use the nonrelativistic potential model for the the description of mesons. 
We perform a short-time expansion in operators of increasing dimensions which we call OPE and 
which has indeed some common features (but also important differences) with the OPE expansion 
in the field theory. We consider

\newpage 
$$\quad\vspace{4.0 cm}$$

\noindent the two cases: regular confining potentials 
\footnote{A regular potential is a potential which is an analytic function of $\vec{r}$ at $r=0$. 
For example, the potential $V(r)\simeq |\vec r|$ falls out of this class.} and singular 
potentials.  

In the SV limit both the amplitude and the decay rate can be formally obtained as a 
{\it double expansion} in $1/m_c$ and $1/\delta m$. 
We consider lowest orders in $1/m_c$, up to $1/m_c^2$, and {\it all orders in $1/\delta m$}. 
Note that this involves terms of much higher order than usually done when one expands in $1/m_Q$ 
with $m_b/m_c$, or as well $\delta m/m_Q$, fixed. Our double expansion allows on the contrary 
to go much further in $1/\delta m$, and this might allow to display subtle duality violations.   

For the regular potential we obtain the full $1/\delta m$ expansion, 
which is only asymptotic to the physical width expanded to the same order in 
$1/m_c$. 
The difference \footnote{As we shall see this expansion contains only a finite number of non zero terms.}
is of order $\delta m/m_c^2\exp(-\delta m/\Lambda)$, which means 
exponentially small duality violation. 

For the singular potential we do not obtain the full $1/\delta m$ expansion: 
following the same procedure as for the regular potential leads to infinite 
coefficients beyond some order in $1/\delta m$. In this case, we find that 
the truncated expansion satisfies duality up to this order.

We consider the inclusive semileptonic decay $B\to X_cl\nu$ in the Shifman-Voloshin (SV) 
limit
$\Lambda\ll\delta m=m_b-m_c\ll m_c,m_b$
and treat mesons as nonrelativistic bound states of spinless quarks in a 
confining potential (a detailed calculation is given in \cite{ourprd}).
This model maximally simplifies both 
constructing the OPE series and calculating the sum of the exclusive channels. 
For the sake of argument we consider the case of leptons coupled to hadrons through 
the scalar current.  
In this case the leptonic tensor is reduced to a scalar function $L(q^2)$.  
The amplitude $T$ depends on the two variables, and we choose them as $q_0$ and $\vec q^2$ in the 
$B$-rest frame:

  
\begin{eqnarray} \label{tproduct}
T(q_0,\vec q^{\,2})=\frac{1}{i}\int
dx {\rm e}^{-iqx} \langle B|T(J(x),J^+(0))|B \rangle\nonumber
\end{eqnarray}
\begin{eqnarray}
\label{t} 
=\sum_X\frac{|\langle B|J|X(-\vec q)\rangle|^2}{M_B -E_X(-\vec q)-q_0}.         
\end{eqnarray}
The sum in (\ref{t}) runs over all hadron states with the appropriate quantum numbers. 
The states are normalized as follows 
$\langle \vec p|\vec p' \rangle =(2\pi)^3 \delta(\vec p-\vec p')$, and  
$E_X(-\vec q)$ is the energy of the state $X$ with the total 3-momentum $-\vec q$. 

At fixed $\vec q^2$, $T(q_0,\vec q^2)$ has a cut in the complex 
$q_0$-plane along the real axis for $q_0<M_B-M_D-\vec q^2/2(m_c+m_d)$, see Fig 1. 

\vskip 0.5 cm
\begin{figure}[hbt] 
\begin{center} 
\mbox{\epsfig{file=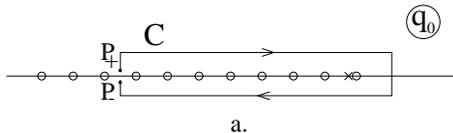,width=6cm}}
\caption{\label{fig:fig1} 
Singularities of the amplitude $T(q_0,\vec q^{\,2})$ in
the complex $q_0$-plane. Circles are poles, corresponding to low-lying charm 
states, and the cross marks the location of the pole in the 
free $b\to c$ quark transition.} 
\end{center}
\end{figure}
\vskip 0.5 cm

A part of this cut for 
$|\vec q|<q_0<M_B-M_D-\vec q^2/2(m_c+m_d)$ corresponds to the decay process. 
The decay rate can be represented as the contour
integral in the complex $q_0$-plane over the contour $C(\vec q^{\,2})$ (Fig 1)  
\begin {eqnarray} 
\label{ratecontour}
\Gamma(B\to X_c l\nu)&=&
\int d\vec q^{\,2}  |\vec q|
\int_{C(\vec q^{\,2})}\frac{dq_0}{2\pi i}L(q^2) 
T(q_0,\vec q^{\,2}).    
\end{eqnarray}
The contour $C(\vec q^2)$ selects at any given $\vec q^{\,2}$
only states kinematically allowed in the decay $B\to X_c l\nu$. 
It is tightly attached to the points $P_{\pm}$ with the coordinates 
($|\vec q|, \pm i0$), otherwise it can be freely deformed in the region where the function 
$T_0(q_0,\vec q^2)$ is analytic. 

The amplitude can be expanded in a series 
\begin{eqnarray} 
\label{ope} 
T(q_0,\vec q^{\,2})=\sum_i c_i(q_0,\vec q^{\,2})\langle B|\hat {\cal O}_i|B \rangle, 
\end{eqnarray}
where $\hat {\cal O}_i$ are  operators  
of increasing dimensions and $c_i(q_0,\vec q^{\,2})$ are
the $c$-number coefficients. Introducing the expansion (\ref{ope}) into
(\ref{ratecontour}) gives the integrated rate as an OPE series\footnote{
The OPE series in the potential model has an important distinctions 
from the Wilsonian scheme in the field theory where contributions 
of distances below the scale $1/\mu$ are referred to the Wilson coefficients, while 
contributions of distances above this scale are referred to the matrix elements. 
As a result, both the Wilson coefficients and the matrix 
elements acquire the $\mu$-dependence. In the potential model, the $T$-product of the two 
currents is also expanded in a series of operators of increasing dimensions, but the 
resulting $c$-number coefficents and the matrix elements in Eq. (\ref{ope}) are scale independent.}

\section{\bf The model}
\vskip.2cm 
Let us proceed along the lines of ref. \cite{ourprd}.
We treat the leptonic part relativistically, but for the description of 
mesons as bound states of spinless quarks use the nonrelativistic potential 
model with a confining potential. We consider the decay in the $B$-rest frame. 
The Hamiltonian of the $b\bar q$ system at rest has the form 

\noindent 
$\widehat H_{bd}=m_b+m_d+\hat h_{bd}$, 
$\hat h_{bd}=\vec k^{\,2}/2m_b+\vec k^{\,2}/2m_d+V_{bd}(r)$, 
such that 
$(\hat h_{bd}-\epsilon_B)|B \rangle=0$, 
$(\hat H_{bd}-M_B)|B \rangle=0$, and $M_B=m_b+m_d+\epsilon_B$. 

The Hamiltonian of the $c\bar q$ system produced in the semileptonic 
$b\to cl\nu$ decay reads 

\noindent
$\widehat H_{cd}(\vec q)=m_c+m_d+(\vec k+\vec q)^2/2m_c+\vec k^{\,2}/2m_d+V_{cd}(r)$. 
The eigenstates of this hamiltonian are $|D_n(\vec q)\rangle$ such that   
$(\widehat H_{cd}(\vec q)-E_{D_n}(\vec q))|D_n(\vec q)\rangle=0$, 
where $E_{D_n}(\vec q)=M_{D_n}+{\vec q^{\,2}}/{2(m_c+m_d)}$ and 
$M_{D_n}=m_c+m_d+\epsilon_{D_n}$. 

The $Q\bar q$ potential can be expanded as follows: 

\centerline{$V_{Qq}=V_0+V_1/{2m_Q}+V_2/{2m_Q^2}+\dots$}
\vskip.4cm
\section{\bf Sum Rules}
\vskip.2cm
The relationship between the sum over the individual channels and the 
meson matrix elements of the operators is established by the sum rules. 
Let us introduce $\delta_n(\vec q)$ through the relation 
($\delta_n(\vec q)=\epsilon_{D_n}-\epsilon_{B}-\vec q^2m_d/2m_c(m_c+m_d)$)
\begin {eqnarray}
\label{deltan}
M_B-q_0-E_{D_n}(\vec q)=\delta m-q_0-{\vec q^2}/{2m_c}-\delta_n(\vec q).
\end{eqnarray}
The $\delta_n(\vec q)$ is the eigenvalue of the operator $\delta H(\vec q)$ 
\begin {eqnarray}
\label{deltah}
M_B-\widehat H_{cd}(\vec q)=\delta m-{\vec q^2}/{2m_c}-\delta H(\vec q)
\end{eqnarray}
with $|D_n(\vec q)\rangle$ the corresponding eigenstates. 
The sum rules are obtained by inserting the full system of the eigenstates 
$|D_n(\vec q)\rangle$ into $\langle B|\left({\delta H(\vec q)}\right)^i|B\rangle$: 
\begin {eqnarray}
\label{sr}
\langle B|\left(\delta H(\vec q)\right)^i|B\rangle=\sum_{n=0}^{\infty}
|F_n(\vec q)|^2 \left(\delta_n(\vec q)\right)^i. 
\end{eqnarray}
where $F_{n}(\vec q)=\langle B|D_n(\vec q)\rangle$ is the $B\to D_n$ transition form factor.  
This relation represents the sum over all $c\bar d$ resonances in terms of the 
$B$-meson matrix element of the operators $\left({\delta H(\vec q)}\right)^i$. 
For the potential regular at the origin $r=0$ the sum over $n$ is convergent for any $i$, 
whereas for the singular potential both sides of Eq. (\ref{sr}) are convergent for small $i$ and 
diverge for large $i$. At the moment we proceed formally and discuss this problem in more detail 
in section E. 

\vskip.4cm
\section{\bf Duality relation for the amplitude}
\vskip.2cm
Making use of the sum rules (\ref{sr}), we represent the amplitude as a sum of the operators: 
\begin{eqnarray}
\label{t1}
T(q_0,\vec q)&=&\sum_{n=0}^{\infty}\frac{|F_n(\vec q)|^2}{M_B-q_0-E_n(\vec q)}
\\
\label{t2}
&=&\frac{1}{\delta m -\frac{\vec q^2}{2m_c}-q_0}
\sum_{n=0}\sum_{i=0}\frac{|F_n(\vec q)|^2(\delta_n(\vec q))^i}
{(\delta m -\frac{\vec q^2}{2m_c}-q_0)^i}
\\
\label{t3}
&=&\frac{1}{\delta m -\frac{\vec q^2}{2m_c}-q_0}
\sum_{i=0}\frac{\langle B|(\delta H(\vec q))^i|\rangle B}{(\delta m -\frac{\vec q^2}{2m_c}-q_0)^i}.
\end{eqnarray}
This expression is the duality relation for the amplitude: the sum (\ref{t1}) runs over the 
infinite number of the charm resonances, and the sum (\ref{t2}) runs over the infinite number of the 
operators of the increasing dimensions (the OPE series). In fact, the location of singularities 
in the complex $q_0$-plane in the series (\ref{t1}) and (\ref{t3}) is quite different: 
in (\ref{t1}) it is an infinite set of single poles at the different locations corresponding to different charm 
resonances, and in (\ref{t2}) it is an infinite set of poles of the increasing order at the same point.  

However this set of equations is only a formal one ; in fact, 
(\ref{t1}) is a summable series leading to a finite result in all cases ; on the other hand, the situation
of eq (\ref{t3}) is more subtle. In the singular case, the coefficients are infinite beyond some order,
and one must accordingly truncate the series. In the regular case, 
the eq (\ref{t3}) is only an asymptotic series : notice that the geometric sum over $i$ in eq (\ref{t2}) has a domain of convergence which is repelled 
to infinity with $n$.

Let us illustrate it with a simple example:  
Assume that $F_n^2\simeq e^{-n}$ and $E_n\simeq n$. Then the analog of the above
equations takes the form 
\begin{eqnarray}
\label{t4}
\sum_{n=0}^{\infty}\frac{e^{-n}}{z-n}=
\sum_{n=0}\frac{e^{-n}}{z} \sum_{i=0}\left(\frac{n}{z}\right)^i
\simeq \frac{1}{z} \sum_{i=0}\frac{i!}{z^i}.
\end{eqnarray}
The last step is obtained by changing the order of summation and using the
relation $\sum_{n=0}^{\infty}e^{-n}n^i\simeq i!$ The series (\ref{t4}) in $i$ is only
asymptotic and not even Borel summable. 

From the amplitude $T$ under the form eq. (\ref{t1}) or eq. (\ref{t3}), respectively, by integration 
over the same contour C, we can obtain either the width as
a sum over the exclusive final states, or as the OPE series. The expression (\ref{t3}) is an accurate
approximation to (\ref{t1}) only when $q_0$ is far from the singularities of $T(q_0,\vec q)$. The contour
C can be deformed away from the singularities except near its fixed end points. When integrating over $q_0$
this is a possible source of discrepancy, i.e. of duality violation.
Consequently, we are now going to estimate the integral of expression (\ref{t1}), i.e. the sum over the
exclusive channels, and the integral of expression (\ref{t3}), i.e. the OPE prediction, and 
compare both results.

\vskip.4cm
\section{\bf The OPE calculation of the decay rate}
\vskip.2cm

Let us first proceed with the amplitude in the form (\ref{t3}) and obtain the OPE 
expression for the decay rate. 
We consider the leptonic tensor of the general form $L(q^2)=(q^2)^N$. For technical reasons, it 
is convenient to isolate $h_{bd}$ in the expression for 
$\delta H(\vec q)$ as follows
\begin{eqnarray}
\label{deltah1}
\delta H(\vec q)&=&h_{bd}-\epsilon_B+\frac{\vec k\vec q}{m_c}
+\left(\frac{1}{m_c}-\frac{1}{m_b}\right)\frac{\vec k^2+V_1}{2}
\end{eqnarray}
Substituting (\ref{deltah1}) in (\ref{t3}) and performing the necessary integrations gives 
a series in $1/m_c$ \cite{ourprd} 
\begin{eqnarray}
\label{gamma-ope}
\frac{\Gamma^{OPE}(B\to X_cl\nu)}{\Gamma(b\to cl\nu)}&=&1+
\frac{\langle B|\vec k^{\,2}|B \rangle}{2m_c^2}
\nonumber\\
&&-(2N+3)\frac{\langle B|V_1|B \rangle}{2m_c^2}
\nonumber\\
&&
+\sum\limits_{i=1}^{2N+3}(-1)^i\frac{C^{i+2}_{2N+5}}{2N+5}\,
\frac{\langle B|\hat O_i|B \rangle}
{m_c^2\delta m^{i}}
\nonumber\\
&&
+O\left(\frac{\Lambda^2\delta m}{m_c^3}\right), 
\end{eqnarray}
with $C^i_{n}=\frac{n!}{i!(n-i)!}$ and $\hat O_i=\vec k(h_{bd}-\epsilon_B)^{i}\vec k$. 
An important feature of the OPE series (\ref{gamma-ope}) is that 
the leading-order term reproduces the free-quark decay rate,  
and the first correction emerges only in the $1/m_c^2$ order
(cf. \cite{cgg,bsuv93}). 

\vskip.4cm
\section{\bf Summation of the exclusive channels}
\vskip.2cm
Now let us sum the rates of the exclusive channels. 
The $B\to D_n$ transition form factors have the form \cite{ourprd} 
\begin{eqnarray}
\label{rho}
F_0^2(\vec q)&=&1-\rho^2_{0}{\vec q^{\,2}}/{m_c^2} 
+O(\vec q^4/m_c^4)
+O({\delta m^2\beta^2}/{m_c^4}),\nonumber\\
F_n^2(\vec q)&=&\quad\;\; \rho^2_{n}{\vec q^{\,2}}/{m_c^2} 
+O(\vec q^4/m_c^4)
+O({\delta m^2\beta^2}/{m_c^4}). 
\nonumber
\end{eqnarray}
Since $|\vec q|\lesssim\delta m$ in the decay region, these expressions allow calculating 
the decay rate to the accuracy $\delta m^2/m_c^2$. Explicitly, we obtain \cite{ourprd}:  
\begin{eqnarray} 
\frac{\Gamma(B\to D_0 l\nu)}{\Gamma(b\to c l\nu)}&=&
1-\frac{3\,\rho_0^2}{2N+5}\frac{\delta m^2}{m_c^2}
+\frac{3}{2}\,\frac{m_d}{1+m_d/m_c}\,\frac{\delta m}{m_c^2}
\nonumber\\
&-&(2N+3)\frac{<B|\vec k^2+V_1|B>}{2m_c^2}, 
\nonumber\\ 
\frac{\Gamma(B\to D_n l\nu)}{\Gamma(b\to c l\nu)}&=&
\frac{3\,\rho_n^2}{2N+5}\,\frac{\delta m^2}{m_c^2} -
3\left(\rho_n^2\Delta_n\right)\frac{\delta m}{m_c^2}
\nonumber\\
&+&\frac{1}{m_c^2}\sum_{i=2}^{2N+5}\frac{(-1)^iC^i_{2N+5}}{2N+5}
\frac{\left(3\rho_n^2\Delta_n^i\right)}{\delta m^{i-2}} ,  
\end{eqnarray}
where $\Delta_n=\epsilon_{D_n}-\epsilon_{D_0}$. 

The main contribution is given by the $B \to D_0$ transition. 
Excited states contribute only starting from the 
$(\delta m)^2/m_c^2$ order in the SV limit. 
Notice that {\it each of the exclusive rates contains terms 
of the order $\delta m^2/m_c^2$ and $\Lambda\delta m/m_c^2$ which are absent in the 
OPE series}. 

Summing over all opened exclusive channels gives 
\begin{eqnarray}
\label{excl}
&&\frac{\Gamma(B\to X_c l\nu)}{\Gamma(b\to c l\nu)}=
1-\frac{\delta m^2}{m_c^2}\frac{3}{2N+5}\left(\rho_0^2-\sum_{n=1}^{n_{max}}\rho_n^2\right)
\nonumber\\
&&\qquad\qquad+3\frac{\delta m}{m_c^2}
\left(\frac{m_d/2}{1+m_d/m_c}-\sum_{n=1}^{n_{max}}\rho_n^2\Delta_n\right)
\nonumber\\
&&\qquad\qquad-(2N+3)\times\frac{\langle B|\vec k^{\,2}+V_1|B\rangle}{2m_c^2}
\nonumber\\
&&\qquad\qquad+\sum_{i=2}^{2N+5}\frac{(-1)^iC^i_{2N+5}}{2N+5}
\frac{3\left(\sum\limits_{n=1}^{n_{max}}\rho_n^2\Delta_n^i\right)}{m_c^2\delta m^{i-2}}. 
\end{eqnarray}
The sum over the charm resonances is
truncated at $n_{max}$, which is the total number of the resonance levels opened at 
$q^2=0$. For the confining potential and in the SV limit $n_{max}$ 
is found from the relation $\Delta_{n_{max}}\simeq \delta m$. 

\section{Check of duality for regular potentials}

The transition radii in the expression (\ref{excl}) are not independent and related to each other 
through the sum rules. These sum rules can be obtained from (\ref{sr}). 
Expanding both sides of (\ref{sr}) in powers of $1/m_Q$ 
and taking the linear $\vec q^2$ term gives the set of the sum rules \cite{ourprd}: 
For $i=0$ one finds the Bjorken sum rule \cite{bjorken}, for $i=1$ - 
the Voloshin sum rule \cite{voloshin}, for $i\ge 2$ - higher moment sum rules:
\begin{eqnarray}
i&=&0:\sum_{n=1}^{\infty}\rho_n^2=\rho_0^2, \qquad 
\nonumber\\
i&=&1:\sum_{n=1}^{\infty}\rho_n^2\Delta_n=\frac{m_d/2}{1+{m_d/m_c}},  
\nonumber\\
\label{sr2}
i&\ge& 2:\sum_{n=1}^{\infty}\rho_n^2\Delta_n^i=
\frac{1}{3}\langle B|\vec k(h_{bd}-\epsilon_B)^{i-2}\vec k|B \rangle. 
\end{eqnarray}

Using these relations to rewrite the OPE result (\ref{gamma-ope}) 
as the sum over hadronic resonances, the difference between the OPE and the
exclusive sum (the duality-violating contribution) explicitly reads 
\begin{eqnarray}
\label{dv}
\delta_{\Gamma}&&\equiv\frac{\Gamma^{OPE}(B\to X_cl\nu)-\Gamma(B\to X_cl\nu)}{\Gamma(b\to cl\nu)}
\nonumber\\
&=&3 \frac{\delta m^2}{m_c^2}\times\sum_{i=0}^{2N+5}\frac{(-1)^iC^{i}_{2N+5}}{(2N+5)\delta m^i}
\sum_{n>n_{max}}\rho_n^2 (\Delta_n)^i
\nonumber\\
&&\qquad+O\left(\frac{\Lambda^2\delta m}{m_c^3}\right) 
\nonumber\\
&=&\frac{\delta m^2}{m_c^2}\frac{3 }{2N+5}
\sum_{n>n_{max}}\rho_n^2 \left(1-\frac{\Delta_n}{\delta m}\right)^{2N+5}
\nonumber\\
&&\qquad+O\left(\frac{\Lambda^2\delta m}{m_c^3}\right)  
\end{eqnarray}
Quite remarkably, $\delta_{\Gamma}$ happens to be equal to the sum of the extrapolated widths 
for charm states beyond the kinematical limit. A similar expression is found in $QCD_2$ \cite{b-thooft}.
Clearly, the duality-violating effect is connected with the 
charm states forbidden kinematically in the decay process.
Notice that $\Gamma^{OPE}(B\to X_cl\nu)-\Gamma(B\to X_cl\nu)<0$, 
because $\Delta_n > \delta m$ for $n>n_{max}$, and $2N+5$ is odd.

To estimate the size of the duality-violation effects, the behavior of the {\it transition radii} and
the relation between $\Delta_n$ and $n_{max}$, which will be given by the behaviour of the  
{\it excitation energies} also at large $n$, are needed. 
For quite a general form of the confining potential we can write the 
following relations for $\Delta_n$ at large $n$ 
$\Delta_{n}\ge\Lambda C n^a$ for $n>n_{max}$ and 
$\Delta_{n_{max}}=\Lambda C (n_{max})^a\simeq \delta m$, 
with $C$ and $a$ some positive numbers. 
In particular, this estimate is valid for the confining potentials 
with a power behavior at large $r$. 
This estimate for $\Delta_n$ is only depending on the behaviour of the potential at large distances.  

The behavior of the radii $\rho_n^2$ at large $n$ are then connected with 
the finiteness of the r.h.s. of the sum rules (\ref{sr2}):   
for a potential regular at $r=0$, the matrix elements in the r.h.s. of the sum rules 
are finite for any $i$, which means that the radii $\rho_n^2$ are decreasing with $n$ faster than any power. 
Essentially this means that $\rho_n^2\simeq \exp(-n)$, and therefore the duality-violating effect in the 
decay rate in (\ref{dv}) is 
of order $\delta_{\Gamma}\simeq\delta m^2/m_c^2 \exp(-\delta m/\Lambda)$. One of such examples, 
the harmonic oscillator potential, is discussed in \cite{ourho}.

\section{Singular potentials}
However, if the potential is singular at $r=0$, the situation changes dramatically.  First, only a few
first number of the matrix elements $\langle B|\hat O_i |B\rangle$ are finite. \footnote{ The appearance
of infinite coefficients in the OPE series is probably due to a breakdown  of the power series expansion,
for instance by fractional  powers or logarithms of $m_Q$ as seems to be the case in the pure Coulomb
case\cite{JC}. For similar phenomena in a perturbation expansion, see \cite{ann}.}

We can try to proceed along the same lines but then have to truncate the series in $1/\delta m$ at the
last finite term. We want to estimate the difference between this truncated series and the exclusive 
sum. 

Let us illustrate this considering a potential 
with a Coulomb behavior at small $r$, $V\simeq -\alpha/r$, and confining at large $r$. Then 
$\langle B|\vec k(h_{bd}-\epsilon_{B})^i \vec k|B\rangle$ are finite for $i\le 1$, but diverge starting 
from $i=2$. We then find that 
\begin{eqnarray}
\rho_n^2\lesssim \frac{1}{n^{1+\varepsilon}}\left(\frac{1}{n^a}\right)^{3}. 
\end{eqnarray}
Such $\rho_n^2$ lead to the estimate  
\begin{eqnarray}
\delta_{\Gamma}\simeq \frac{\Lambda^2}{m_c^2}\left(\frac{\Lambda}{\delta m}\right)^{(1+\varepsilon/a)}.   
\end{eqnarray}

More generally, if the above matrix element begins to diverge for some value $i=K+1$, the formulas are to be
replaced by :
\begin{eqnarray}
\rho_n^2\lesssim \frac{1}{n^{1+\varepsilon}}\left(\frac{1}{n^a}\right)^{K+2}. 
\end{eqnarray}
 
\begin{eqnarray}
\delta_{\Gamma}\simeq \frac{\Lambda^2}{m_c^2}\left(\frac{\Lambda}{\delta m}\right)^{(K+\varepsilon/a)}.   
\end{eqnarray}

Notice that this $\delta_{\Gamma}$ is smaller than the last retained term in the OPE series which is 
of order $\frac{\Lambda^2}{m_c^2}\left(\frac{\Lambda}{\delta m}\right)^{K}$. 
Therefore the 'duality violation' is just smaller than the last retained term as for the asymptotic series.
This means in fact that there is {\it no indication of duality violation at this computable order}. This is
independent of $a$, therefore of the {\it large distance behavior of the potential}.   

\section{Conclusion}

Summarizing our results, the amplitude $T(q_0,\vec q^2)$ - the T-product, eq. (\ref{tproduct}) -
can be expanded in inverse powers of
$\delta m-\vec q^2/2m_c-q_0$, the so-called OPE expansion. Exact duality would mean that the OPE
series was convergent and equal to $T$. Actually, this is not exactly the case. Even in the favourable
case of the {\it regular potentials} (at $\vec r=0$), the OPE series is not convergent,  it is only
asymptotic to the actual $T$. For {\it singular potentials}, the coefficients are simply infinite beyond a
certain order.

Besides these problems concerning the amplitude, additional problems appear for the expansion of the width,
which is given by a contour integral of $T$ in the $q_0$ complex plane : the OPE expansion is accurate 
far from the singularities in $q_0$, while the contour has fixed end points in the complex plane close 
to the singularities (Fig. 1). In view of this situation, we have computed explicitly the difference 
between the OPE and the actual width. For singular potentials, the series must be truncated, and the 
difference is found smaller than the last retained term.

As to the perspectives opened by this work, we must first emphasize that singular potentials seem
more interesting than regular ones. 
Indeed, in QCD the effective quark potential is singular, a smoothed
Coulomb singularity. Moreover, in $QCD_2$, one can suspect some similarity with a linear potential 
$|\vec r|$, which is also singular at the origin in the sense of this paper. 
For a singular potential, we have seen that the entire series must be truncated at some order, because
the coefficients become eventually infinite.
We think that such infinite coefficients in an entire series expansion correspond to the fact that the
correct expansion is not entire but must include fractional powers and/or logarithms in the expansion
parameter, i.e. $\delta m$.
In QCD, one can argue that the operator matrix elements are finite due to renormalisation, but 
nevertheless the coefficients still contain logarithms of heavy masses. In the non-relativistic case, the
object of the present paper, the method which has been followed does not lead to definite conclusions as
regards duality for singular potentials : namely, to the order we are able to calculate in this paper, we
find that there is no duality violation, but this leaves open the question of duality violation at some
higher order  \footnote{In the context of QCD$_2$, one has demonstrated duality up to the order $1/m_Q^4$
and it may be believed  that duality has been fully demonstrated in higher orders \cite{b-thooft}.
However, a comment is in order here. In \cite{b-thooft}, it was shown that the matrix element of the 
leading operator $\langle B|\bar Q Q|B\rangle$ is dual to the sum of the widths of the full tower of
resonances. Therefore, one can suspect that there is a difference between the actual width and the OPE,
that is of higher order $1/m_Q^5$, corresponding to the extrapolated width of the kinematically forbidden
states. This difference, however, has the same order $1/m_Q^5$ as the matrix elements of the higher
dimension operators \cite{b-thooft}.   It was then {\it assumed} that both quantities are dual to each
other, but the corresponding  OPE coefficients were not calculated and we have not found where this
assumption was demonstrated.}.  To proceed further, one would have to devise new methods to obtain the
above conjectured generalized  expansions.

We are grateful to G. Korchemsky, B. Stech for discussions, and specially to N. Uraltsev for detailed
correspondence. 
D.M. was supported by BMBF under project 05 HT 9 HVA3. 
Laboratoire de Physique Th\'eorique is Unit\'e Mixte de Recherche CNRS-UMR8627. 

\end{document}